\documentclass[11pt]{article}
\usepackage{amssymb}
\usepackage{amsmath}
\newtheorem{theorem}{Theorem}
\newtheorem{lemma}{Lemma}

\newtheorem{definition}{Definition}
\newtheorem{remark}{Remark}
\newtheorem{deff}{Definition}

\newtheorem{open}{Open Question}

\newtheorem{claim}{Claim}
\newtheorem{thm}{Theorem}

\newcommand{\wbox} {\mbox {$\sqcap$\llap {$\sqcup$}}}

\newcommand{\ignore}[1]{}
\newcommand{\ket}[1]{|#1\rangle}
\newcommand{\bra}[1]{\langle#1|}

\newcommand{\beq}[1]{\begin{equation}}
\newcommand{\enq}[0]{\end{equation}}

\newcommand{\ra}{\rangle}
\newenvironment{proof}{{\noindent{\bf Proof:}}}{$\Box$}
\newcommand{\la}{\langle}

\title{Quantum {\cal NP} - A Survey}
\author{Dorit Aharonov\thanks{School of Engineering and Computer Science, The Hebrew University, Jerusalem, Israel, and the Mathematical Sciences Research Institute, Berkeley, California} and Tomer Naveh\thanks{Department for Computer Science, Tel Aviv University, Tel Aviv, Israel}}
\date{}

\begin{document}
\pagestyle{plain}
\pagenumbering{arabic}

\maketitle
\begin{abstract}
We describe Kitaev's result from 1999, in which he defines the complexity 
class {\cal QMA}, the quantum analog of the class NP, and shows that 
a natural extension of $3-$SAT, namely local Hamiltonians, is {\cal QMA} 
complete. The result builds upon the classical Cook-Levin proof of the
NP completeness of $SAT$, but differs from it in several fundamental ways,
 which we highlight. This result raises a rich array of
 open problems related to quantum complexity, algorithms and entanglement, 
which we state at the end of this survey.
This survey is the extension of lecture notes taken by Naveh 
for Aharonov's quantum computation course, held in Tel Aviv University, 2001.
\end{abstract}

\section{Introduction}
The field of complexity theory has witnessed several fundamental results 
in the recent decade or two; It is now a rich field involving 
deep questions and leading to the discovery of 
 beautiful and unexpected structures, with important contributions to 
the understanding of the notion of classical probabilistic and 
deterministic computation. 
With the stormy entrance of quantum computation into the 
life of theoretical computer scientists, it seems only natural to ask 
whether such a rich theory of complexity can also be developed
for the quantum model; it is probably true that such interesting structures 
and results await for us down the road of quantum complexity theory, 
with perhaps insights to be drawn from them regarding the 
quantum computational power. Several important results have already 
been discovered\cite{watrous, kitwat}, and there are surely more to come. 
 It is not unreasonable to also hope that quantum complexity can  
significantly contribute to the understanding of classical complexity
in unexpected ways; A puzzling example in which quantum 
arguments are used in order to prove 
an entirely classical result in the area of locally decodable codes
 was recently found\cite{jordan}.
 We thus view the development of the 
field of quantum complexity as an important
direction that holds the promise of a rich area of study
with possible implications to the understanding of quantum algorithmic theory, 
as well as to classical complexity theory and 
to the foundations of quantum physics. 

Perhaps the most basic and fundamental 
 result in classical complexity theory, is the
 Cook-Levin theorem\cite{cooklevin}, which
states that $SAT$, the problem of whether a Boolean formula is 
satisfiable or not, is $NP$ complete. 
This result  opened the door to the study of the extremely expressive 
complexity class $NP$, and the rich theory of $NP$-completeness, 
and was an important building block in many later 
results in theoretical computer science and complexity theory, 
such as the PCP theorems, hardness of approximation results and the 
proof that $IP=PSpace$. In the heart of this result stands the very 
basic understanding that computation is local.

We devote this manuscript to the survey of a result by Kitaev\cite{kitaev, kitaevbook}, 
which is the quantum analog of the Cook-Levin Theorem. 
Kitaev first defines the quantum analog of $NP$, and then defines 
a complete problem which can be viewed as a generalization of $SAT$ 
to the quantum world. The proof follows the lines of the 
Cook-Levin proof, but defers from it in some fundamental points; 
We highlight those as we go along.
The classical proof is quite simple; The quantum counterpart is rather 
complicated and long.  However, there are several reasons to study
 this theorem,  apart from the elegance of the proof, and from the naturalness 
of the question. First, there is a lot to be learned
 from the comparison of the 
classical proof and its (much more involved) quantum
 counterpart; Understanding the exact places where those differ is 
insightful. Secondly,  the result raises a rich array of 
natural and interesting open problems related to this subject; 
We list those at the end of the survey, after the proof.
Our proof follows closely the proof given by Kitaev\cite{kitaev, kitaevbook},
 with minor 
deviations; Our main contribution here is adding explanations 
and clarifications, hopefully providing some intuition behind the 
proof, and highlighting some open problems. 
We hope that this survey will provide an easy access 
to Kitaev's fundamental result and to the rich array of open questions 
 it raises. 

\section{Definition of {\cal QMA}}
We would like to define a complexity class which will be the quantum 
analog of NP:
\begin{definition}{\bf NP:}
\label{d1} $L \in NP$ if there exists a deterministic polynomial time 
verifier $V$ such that:
\begin{itemize}
\item{$\forall x \in L ~~ \exists y ~~ |y|=poly(|x|), ~ V(x,y)=1.$}
\item{$\forall x \notin L ~~ \forall y ~~ |y|=poly(|x|), ~ V(x,y)=0.$}
\end{itemize}
\end{definition}

By $|x|$ we mean the number of bits in the binary string $x$. 
However, when trying to define the quantum analog, 
 we immediately encounter an 
obstacle. We cannot require the verifier to answer $0$ 
or $1$ deterministically, because we will not be able 
to distinguish between this case and the case in which 
the verifier outputs these values with extremely high 
probability. Since the fact that states are continuous is inherent 
 to quantum computation, we resort to defining {\cal QMA}, the quantum analog of {\cal MA}, 
which is the probabilistic version of {\cal NP}. 

Informally, {\cal MA} can be thought of as a probabilistic analog of {\cal NP}, 
allowing for two-sided errors.
\begin{definition}\underline{{\cal MA:}} ~~
\label{d1} $L \in {\cal MA}$ if there exists a probabilistic polynomial time verifier $V$ such that:
\begin{itemize}
\item{$\forall x \in L ~~ \exists y ~~ |y|=poly(|x|), ~ Pr(V(x,y)=1) \geq 
\frac{2}{3}$}
\item{$\forall x \notin L ~~ \forall y ~~ |y|=poly(|x|), ~ 
Pr(V(x,y)=1) \leq \frac{1}{3}$}
\end{itemize}
\end{definition}

{\cal MA} is naturally viewed as a game or interaction between 2 parties - 
Merlin, which has infinite computational power, and Arthur, 
which is limited to a polynomial time machine (the above $V$). 
Merlin should answer queries such as ``is $x \in L$?'', 
and accompany the answer with a polynomial witness $y$ which 
Arthur can verify in polynomial time. Note that when showing that a problem 
is in {\cal MA}, we should also show that Merlin cannot fool Arthur - 
i.e. that when $x \notin L$ there is no witness $y$ that can 
persuade the verifier to believe that $x \in L$ with probability 
$\ge \frac{1}{3}$.

We will define {\cal QMA} analogously, where the verifier $V$ is a 
quantum machine, and the witness $y$ is a state of a polynomial 
number of qubits. 
We denote by ${\cal B}$ the Hilbert space of one qubit.

\begin{definition} \underline{{\cal QMA}:} ~~
\label{dd1} $L \in {\cal QMA}$ if there exists a quantum polynomial time verifier $V$ and a polynomial $p$ such that:
\begin{itemize}
\item{$\forall x \in L ~\exists |\xi\ra \in {\cal B}^{p(|x|)}, ~ Pr(V(|x\ra|\xi\ra)=1) \geq 
2/3$}
\item{$\forall x \notin L ~~ \forall |\xi\ra \in {\cal B}^{p(|x|)},~
Pr(V(|x\ra|\xi\ra)=1) \leq 1/3$}
\end{itemize}
\end{definition}

Another possible definition would be to 
take $|\alpha\ra$ as a classical witness, i.e. a basis state, but leave 
$V$ to be a quantum  machine. We call this class Quantum Classical {\cal MA} ({\cal QCMA}).

\begin{definition} \underline{{\cal QCMA:}}~~
\label{dd1} $L \in$ {\cal QCMA} if there exists a quantum polynomial time verifier $V$ and a polynomial $p$ such that:
\begin{itemize}
\item{$\forall x \in L ~~ ~\exists y ~~ |y|=poly(|x|), ~ Pr(V(|x\ra|y\ra)=1) \geq 2/3$}
\item{$\forall x \notin L ~~ \forall y, ~~ |y|=poly(|x|), ~
Pr(V(|x\ra|\alpha\ra)=1) \leq 1/3$}
\end{itemize}
\end{definition}

\begin{claim}
 {\cal MA} $\subseteq$ {\cal  QCMA}$ \subseteq${\cal  QMA}.
\end{claim}

\begin{proof} The left inclusion is trivial. The right inclusion follows from the fact that the quantum verifier can force Merlin to send him a classical 
witness by measuring the witness before applying on it 
the quantum algorithm.\end{proof}

It is unclear whether the two classes, $QCMA$ and $QMA$ are the same; 
See open question \ref{same} for further discussion. 
In any case, for the purposes of this paper, 
we will limit ourselves to the class {\cal QMA}, where 
the witnesses are quantum.  

\subsection{Amplification}
In all the above definitions of $MA, QCMA$ and $QMA$, 
 we have used as our completeness 
parameter (i.e. one minus the error probability in case $x\in L$)
 the value $2/3$ and as our soundness parameter (the bound on error probability in case $x \not\in L$) the value $1/3$. 
We can denote this choice by {\cal MA}$(2/3,1/3)$ or {\cal QMA}$(2/3,1/3)$. 
In general, we can define the classes {\cal MA}$(c,s)$ or {\cal QMA}$(c,s)$
with general completeness and soundness parameters, which are functions of the 
input's length. It turns out that 
 we have a lot of freedom in the choice of these parameters, 
and we can make them either polynomially close to each other, or 
exponentially close to $1$ or $0$, without changing the complexity classes 
we are dealing with. In other words, amplification of the completeness 
and the soundness from polynomial separation to exponentially  small error 
can be done in polynomial overhead. This is done using parallel repetition  
and taking the appropriate majority. More formally, for the case of the classical class $MA$:  

\begin{thm}
{\cal MA}$(c,c-1/n^g)\subseteq$ {\cal MA}$(2/3,1/3)=${\cal MA}$(1-e^{-n^g},e^{-n^g})$
where we require $g$ to be a constant and $0< c, c-1/n^g< 1$.  
\end{thm}

\begin{proof}
If $c$ and $s$ are separated by some $1/poly(n)$,  we 
run the verifier polynomially many times, say $m$, 
using independent random coins at each time. 
In case $x\in L$ the expected number of acceptances is at least $cm$, whereas 
in case  $x\not\in L$ it is at most $sm$; 
The Chernoff bound\footnote{The Chernoff bound guarantees that the average of 
polynomially many repetitions of independent experiments will 
converge exponentially fast to the expected value}\cite{chernoff}
guarantees that we can distinguish between the cases with only 
polynomially number of independent experiments with exponentially small error. 
This proves the inclusions ${\cal MA}(c,c-1/n^g)\subseteq {\cal MA}(2/3,1/3)\subseteq {\cal MA}(1-e^{-n^g},e^{-n^g})$ where the other direction is trivial. 
\end{proof}
 
Hence, we can conveniently move between the definition of {\cal MA} 
with either one of these three possible choices of parameters.

\begin{remark}  The class {\cal MA} as we defined it has two sided errors; 
In fact, this class is equivalent to  {\cal MA}  with only one sided 
error, i.e. with completeness $1$ and soundness bounded away from $1$\cite{zachusfurer, goldreichzuckerman}. It is unclear whether the same holds in the quantum case; See open question \ref{two}. 
\end{remark}

This nice freedom in the choice of parameters, due to the 
parallel repetition, holds also in the quantum case, however with 
a slightly more complicated proof. 

\begin{thm}
{\cal QMA}$(c,c-1/n^g)\subseteq$ {\cal QMA}$(2/3,1/3)=${\cal QMA}$(1-e^{-n^g},e^{-n^g})$
where we require $g$ to be a constant,   $0< c, c-1/n^g< 1$.
\end{thm}

\begin{proof}
The proof of this theorem is slightly more subtle than the 
simple proof in the classical case. We will first prove that 
 {\cal QMA}$(2/3,1/3)$ is contained in {\cal QMA}$(1-e^{-n^g},e^{-n^g})$,  
i.e. that we can amplify soundness and completeness exponentially. 
The idea of remains the same as in the classical case:
 the verifier 
should perform polynomially many independent experiments
and output the majority vote. 
However, unlike in the classical case, the verifier cannot perform 
many independent experiments on the same witness provided by the prover
 since after measuring it 
the witness will have changed; Neither can the verifier
 copy the quantum witness
 state before verifying it, due to the no cloning theorem\cite{nocloning}
which states that an unknown quantum state cannot be copied.      
  The verifier thus needs to ask the prover 
to provide him with polynomially many copies of the witness. 
This is problematic, since the prover might try to cheat by entangling 
the witnesses he provides. We will have to show
 that such a strategy cannot help 
the prover in case $x$ is not in the language. 

We construct a new verifier which runs in parallel 
polynomially many copies of the verifier $V$, then outputs 
the majority. The existence of a witness for the 
new verifier in case $x\in L$ is trivial since it is simply 
duplicate copies of the original witness. 
To prove soundness, one might suspect that entanglement 
between the provers can be used to bypass the fact that 
the error goes exponentially to $0$. To show this cannot happen, 
we treat the verifiers  as if they are applied
 one after the other, and not in parallel. This is correct 
since the verifiers operate on different qubits and so they commute. 
We know that the probability that the first 
copy of $V$ outputs $1$ is less than $1/3$.
After the first verifier was applied, we can apply the second verifier. 
The second verifier gets as an input some state, which can be conditioned 
on the result of the measurement of the first verifier.  
However, regardless of what this output was, it is still 
correct that the probability for an output $1$ is less than $1/3$.
And so on for the remaining of the verifiers.  
Hence, the probability for the majority of the verifiers being $1$ 
can be bounded from above by the probability that polynomially many 
independent Bernoulli trials with bias $1/3$ will be $1$, which 
  decays exponentially by Chernoff.

The idea of the inclusion {\cal QMA}$(c,c-1/n^g) \subset 
${\cal QMA}$(2/3,1/3)$ 
is exactly the same, except that instead of majority vote among the 
polynomially many verifiers, we need 
to count the number of accepting verifiers, and accept only if this number 
is above $(c+s)/2$ times the number of experiments. 
The other inclusions are trivial.  
\end{proof}

In the rest of the survey, we will interchange between 
the different choices of parameters according to our convenience.

\subsection{Complexity}
Before we continue, let us summarize what is known about these 
classes in terms of complexity.  
The most important class in quantum complexity theory is the
 class $BQP$, which consists
 of those problems which can be solved by a quantum 
machine with error probability bounded below half; This is considered as 
the class of tractable problems on a quantum computer.
It is to be compared with the class $BPP$ which is the same class for 
classical computers. Of course, we have that 
$BPP\subseteq BQP$, and that $BQP\subseteq QMA$. 
But how powerful is the class $QMA$? 
Can we upperbound it? 
 Adleman {\it et. al.} proved that $BQP$ 
is contained in a large class, called $PP$. 
A language $L$ is in $PP$ if there exists a Turing machine 
that runs in polynomial time on an input $x$, and such that if $x\in L$ 
it outputs $1$ with probability larger than $1/2$, 
and if $x\not\in L$ it outputs $0$ with probability larger than $1/2$. 
Note that the difference between the output probability and $1/2$ 
can be exponentially small.
This makes the class possibly much stronger than the class $BPP$;  
In particular, $PP$ contains $NP$ (see \cite{goldreich} lecture $7$).
It turns out that the above upper bound on $BQP$ 
can be generalized
to prove the same inclusions for the class $QMA$, i.e. 
$QMA \subseteq PP$. This fact was first noted by Kitaev and 
Watrous\cite{watrous2} who build on
a simplification of \cite{adleman} by
 Fortnow and Rogers\cite{fortnow} to prove it. 
 To summarize we have that: 

\begin{thm}
$BPP\subseteq BQP\subseteq QCMA \subseteq QMA \subseteq PP.$
\end{thm}

This is almost all that is known regarding the relation of $BQP$ and $QMA$ 
to classical complexity classes.  
To give intuition about what this upper bound
 means regarding the quantum complexity 
power, we note that the class $PP$ is known to be contained in  perhaps
a more natural class, $PSPACE$,    
which is the class of languages that can be recognized by a Turing machine 
that uses polynomial {\it space} (but can take exponential amount of time.) 

We now proceed to define the complete problem for $QMA$: Local Hamiltonian.

\section{The Local Hamiltonian Problem}
In this section we will define what can be thought of as the quantum 
analog of $3-SAT$, called the ``local Hamiltonian problem''.

\begin{deff}{\bf 5-Local Hamiltonian problem}
\begin{itemize}
\item  {\bf Input:} $H_1,...H_r$, A set of $r$ Hermitian 
positive semi definite matrices operating on the space of five qubits, 
${\cal B}^{\otimes 5}$, 
 with bounded norm $\|H_i\|\le 1$. 
Each matrix comes with a specification of the $5$ qubits
(out of the total $n$  
qubits) on which it operates.
 Each matrix entry is given with poly(n) many bits. 
Apart from $H_i$ we are also given
 two real numbers, $a$ and $b$ (again, with polynomially
 many bits)  
such that $b-a>1/poly(n)$.  
\item {\bf Output:} Is the smallest eigenvalue 
of $H=H_1+H_2+...+H_r$ smaller than $a$ or are all eigenvalues 
larger than $b$?  
\end{itemize} 
\end{deff}

We slightly abuse notation here by writing $H=H_1+H_2+...+H_r$; 
$H_i$ are matrices operating on different qubits, and the summation 
is over their extension to the entire set of qubits (tensor product
 with identity). This abuse of notation will be used throughout the paper, 
and it will be clear that we mean the summation of the operators 
as operators on $n$ qubits. 

Note that the defined problem
 is a promise problem: we are promised that one of the two 
possible outputs occurs. In other words, we don't care what 
 the output is for Hamiltonians with minimal eigenvalue between $a$ 
and $b$.  

In the same way, one can naturally
define the $k$-local Hamiltonian problem for any $k$.
We will see that $5$-local Hamiltonian is {\cal QMA} complete, and 
 the reason for the number five will only be apparent towards the very end of the proof. However it is unclear whether it is necessary to consider $5$ 
local Hamiltonian or whether a smaller number suffices; 
 See open question number \ref{more}
for further discussion.

\ignore{the  log local Hamiltonian problem, and it is defined 
as follows.

\begin{deff}
{\bf log local Hamiltonians problem}
This is defined exactly as local Hamiltonians problem, except that 
 instead of qubits, the registers can carry polynomially 
many values, i.e. each register is of dimensionality 
which is bounded by a polynomial in the number of registers. 
The Hamiltonians $H_i$ still operate only on a constant number of registers
(we will use  here  $k=3$.) 
\end{deff}}

\ignore{
For the sake of the proof, we also define a slightly more restricted 
version of these 
problems, which are called local projections problem.
\begin{deff} 
{\bf local and log local Projections problems}
These are defined exactly as the local and log
local Hamiltonians problem, except that
we restrict $H_i$ to be a projection on its qubits or registers,
 i.e. $H_i=|\alpha_i\ra\la \alpha_i|$. 
\end{deff}
}

\subsection{Connection to $3-SAT$}
We now show that the local Hamiltonian problem is a 
natural generalization of $3-SAT$ to the quantum world.
For this, we explain how $3-SAT$ can be viewed as a $3$-local Hamiltonian 
problem. We will work with qubits, but all operations are now classical 
operations in disguise.  
Let $\phi = C_1 \wedge C_2 \wedge \dots \wedge C_r$ be a 3-SAT formula on $n$ 
variables, where each $C_i$ is a clause, i.e. an OR over three variables 
or their negations.
 For every clause $C_i$ we define a $8 \times 8$ matrix $H_i$,
operating on three qubits. $H_i$  
 is a projection on the unsatisfying assignment of $C_i$.   
For example, for the clause $C_i=X_1 \vee X_2 \vee \neg X_3$ we get the matrix:
\begin{displaymath}
H_i=     \left ( \begin{array}{cccccccc}        
                0&0&0&0&0&0&0&0\\
                0&1&0&0&0&0&0&0\\
                0&0&0&0&0&0&0&0\\
                0&0&0&0&0&0&0&0\\
                0&0&0&0&0&0&0&0\\
                0&0&0&0&0&0&0&0\\
                0&0&0&0&0&0&0&0\\
                0&0&0&0&0&0&0&0\\
               \end{array}
\right )=|001\ra\la 001|
\end{displaymath}
since $001$ is the only unsatisfying assignment for $C_i$. 
$H_i$ defined this way is a projection matrix. 
Moreover, it is Hermitian. 
If we look at $H_i$, its eigenvectors are all basis vectors 
of three qubits, with  the vectors corresponding to satisfying assignments 
having eigenvalues $0$, and the vector of the unsatisfying 
assignment corresponding to the eigenvalue $1$.
We then consider the operation of $H_i$ on all the qubits, 
by taking tensor product of $H_i$ with identity on the rest of the qubits. 
We denote the new matrix by $H_i$ too, again by slight abuse of notation; 
It will be clear from context which of these we are talking about.  
 If $z$ is an assignment to the $n$ variables 
which satisfies a clause $C_i$, then $H_i|z\ra = 0$. 
Otherwise,  $H_i |z\ra = |z\ra$.
We can view this as if the matrix $H_i$ ``penalizes''
 assignments that do not satisfy $C_i$ by giving them one unit 
of ``energy''. 
We denote $H = \sum_{i=1}^{r} H_i$, and 
 observe that  $H|z\ra = q|z\ra$ where $q$ 
is the number of clauses unsatisfied by $z$.
All eigenvalues of $H$ are non negative integers, and 
 zero is an eigenvalue of $H$ if and only if $H$ corresponds to
 a satisfiable 
formula. Otherwise, the smallest eigenvalue of $H$ is at least $1$. 
Thus, $3-SAT$ is equivalent to the following problem:
 ``Is the smallest eigenvalue of $H$  $0$ or is it at least $1$?'', 
which is an instance of the $3-$local Hamiltonian problem.

\section{Local Hamiltonians is in {\cal QMA}}

\begin{theorem}
The $k$-Local Hamiltonian problem is in {\cal QMA} for any $k=O(log(n))$. 
\end{theorem}

\begin{proof}
We first want to show that if the Hamiltonian $H$ has
 an eigenvalue smaller than 
$a$, i.e. if we are in a ``yes'' instance,  
then there exists a witness that Marlin can use to convince Arthur
for this fact. The obvious witness to use, is simply an eigenstate
with eigenvalue smaller than $a$. 
Let us denote this ground state by $|\eta\ra$ and its corresponding 
 eigenvalue by $\lambda$. 
We will construct a procedure which outputs $1$ with probability 
which is related to this eigenvalue. 
To illustrate the idea, consider first the simpler case in which 
all the Hamiltonians $H_i$ are merely projections,
 $H_i=|\alpha_i\ra\la\alpha_i|$. 
In this case, we note that 
\begin{equation}
\lambda=\la\eta|H|\eta\ra=
\sum_{i=1}^r \la\eta|H_i|\eta\ra=\sum_{i=1}^r \la\eta|\alpha_i\ra\la\alpha_i|\eta\ra=\sum_{i=1}^r |\la\eta|\alpha_i\ra|^2
\end{equation}
or, 
\begin{equation}\label{inter}
\lambda/r=(1/r)\sum_{i=1}^r |\la\eta|\alpha_i\ra|^2
\end{equation}
We note that $|\la\eta|\alpha_i\ra|^2$ is exactly the probability to get 
a positive answer when measuring the state $|\eta\ra$ in the basis 
$|\alpha_i\ra$ and the subspace orthogonal to it. 
Thus, equation \ref{inter} gives the following interpretation 
of $\lambda/r$: It is simply the probability to get the answer $1$
 when we pick $i$ 
randomly between $1$ and $r$ and measure $|\eta\ra$ in the basis 
$|\alpha_i\ra$ and the subspace orthogonal to it. 
This implies an easy way to design an experiment,
 or a quantum verification procedure on the input state 
$\eta$, which outputs $1$ with probability $1-\lambda/r$ and $0$ otherwise. 
Pick an $i \in \{1,...r\}$ uniformly at random, 
measure $|\eta\ra$ in the basis $|\alpha_i\ra$ and the orthogonal subspace, 
and output $0$ if the measurement resulted in a projection on $|\alpha\ra$; 
output $1$ otherwise. 
The probability for $1$ is exactly $1-\lambda/r\ge 1-a/r$. 
On the other hand, if $H$ is a ``no'' instance, 
i.e. all eigenvalues are larger than $b$, then 
for any vector $|\eta\ra$, 
\begin{equation}
\la\eta|H|\eta\ra=\sum_{i=1}^r \la\eta|H_i|\eta\ra \ge b;
\end{equation}
The probability for $1$ in the experiment 
is in this case 
\begin{equation}
1-\la\eta|H|\eta\ra/r\le 1- b/r;
\end{equation}
Since we know that $b-a\ge 1/n^g$, we also have that 
the probabilities for $1$ for the ``yes'' and ``no'' instances are polynomially different: $1-a/r>1- b/r+1/n^g$. We can amplify this difference 
 using the amplification theorem and this proves that 
the problem is indeed in {\cal QMA}, if the Hamiltonians are simple 
projections. 

We remark that since the projections are local, i.e. involve
 at most $log(n)$ qubits,  
 such a measurement can be performed by a polynomial quantum verifier.

To deal with the more general case, 
where $H_i$ are general Hermitian positive semidefinite matrices with norm 
at most $1$, 
we note that any such matrix can be written in its spectral decomposition, 
\begin{equation}
H_i=\sum_{j=1}^{dim(H_i)} w_j^i |\alpha_j^i\ra \la\alpha_j^i| 
\end{equation}

We now impose the following trick which enables us
 to toss a coin with probability 
$1-\la\eta|H_i|\eta\ra$. 
We first add one qubit to the system, in the state $|0\ra$. 
We then apply the following unitary transformation on the qubits 
of $H_i$ and on the extra qubit:  
\begin{equation}
T|\alpha_j^i\ra|0\ra =|\alpha_j^i\ra(\sqrt{w^i_j}|0\ra+\sqrt{1-w^i_j}|1\ra)
\end{equation}
We now prove that the measurement of the extra qubit 
outputs $1$ with probability  $1-\la\eta|H_i|\eta\ra$. 
To see this, write  
\begin{equation}
|\eta\ra =\sum_j y_j|\alpha_j^i\ra|\beta_j^i\ra
\end{equation}
using the Schmidt decomposition. 
After the transformation $T$, 
this state evolves to 
\begin{equation}
T|\eta\ra =\sum_j y_j|\alpha_j^i\ra|\beta_j^i\ra(\sqrt{w^i_j}|0\ra+\sqrt{1-w^i_j}|1\ra)
\end{equation}
The probability to measure $1$ is then the squared
 norm of the following vector: 
\begin{equation}
\sum_j \sqrt{1-w^i_j}y_j|\alpha_j^i\ra|\beta_j^i\ra.
\end{equation}
This squared norm is just
\begin{equation}
\sum_j (1-w^i_j)|y_j|^2=1-\sum_j w^i_j|y_j|^2
\end{equation}
but we know that 
\begin{equation}
\la\eta|H_i|\eta\ra=\sum_j w^i_j|y_j|^2.
\end{equation}
We can now describe the exact verification procedure: 
Pick a random index $i$, and perform the above test for $H_i$: 
Add one qubit, apply $T$ and measure the extra qubit.  
The outcome will be $1$ with probability 
\begin{equation}
\sum_i (1/r) (1- \la\eta|H_i|\eta\ra)=1- \la\eta|H|\eta\ra/r
\end{equation}
If we are in a ``yes'' instance, this number will be larger 
than $1-a/r$; If we are in a ``no'' instance, it will be smaller 
than $1-b/r$, and the proof is completed just as in the simple 
projections case. \end{proof}

\section{QMA Completeness}
In this section we will show that the $5$-local 
Hamiltonian problem is QMA-hard.
The proof is complicated, and we will start with an overview. 

{~}

\subsection{Reminder of the Cook-Levin Proof}
The proof that local Hamiltonian is QMA complete 
bears a lot of  resemblance to Cook-Levin's 
proof that 3SAT is  NP-Complete.
Let us briefly sketch the idea underlying the
 Cook-Levin's proof, so that we can refer to it later on. 
 Consider an NP problem, $L$. 
There is a Turing machine  which operates on $x,y$ where 
$x$ is a supposedly member of $L$ and $y$ is a supposed witness for this fact, and $M$ checks that $x$ is in the language using $y$. 
We now want to construct a reduction to $3-SAT$, i.e. to design a Boolean 
formula which is satisfiable if and only if $x$ is in the language, i.e. 
if the Turing machine performed a successful computation which 
started with the input and ended with ``accept'' or $1$, in the first 
site on the tape.  To construct such a formula, we consider the
 variables $x_{i,t}$ 
where $i$ runs over all reachable locations on the tape in the polynomial 
time limit and $t$ runs over the time steps. $x_{i,t}$ are variables which 
can get any of some constant number of possible values; These values 
correspond to a finite description of the state of the Turing machine 
related to the location $i$ on the tape at time $t$. They include what 
is written on the tape at that time and that location, the state of the Turing 
machine at that time, and whether the head of the Turing machine 
is at that location or not. An assignment to these variables can be viewed as
a {\it history} of some computation; a description of how the Turing 
machine evolved in time. 
The 3-SAT formula we construct is essentially checking that this 
evolution is a valid evolution of the Turing machine. 
Each clause in the formula will look at three subsequent cells 
at some time $t$, say $x_{i-1,t}, x_{i,t}$ and $x_{i+1,t}$ plus 
the cell  $x_{i,t+1}$. Given the values of 
$x_{i-1,t}, x_{i,t}$ and $x_{i+1,t}$, it is possible to know whether 
the value of $x_{i,t+1}$ is valid or not; Thus, the clause 
is satisfied if and only if $x_{i,t+1}$ evolves from $x_{i-1,t}, x_{i,t}$ and $x_{i+1,t}$ by a valid computation. 
We also add clauses that check that the input is really $x$, 
i.e. clauses of the form $x_{i,0}$ if the $i'th$ input bit was $1$, and 
$\neg x_{i,0}$ if the $i'th$ input bit was $0$. 
Finally, we check  that the output is accept by adding the clause $x_{1,T}$, 
which is satisfied if the first site is $1$ at the end of the computation at 
time $T$. 
Each of these verifications is local, since the evolution of the 
Turing machine is local, and thus each corresponds to a clause. 
Note that our variables have a constant but possibly large 
set of possible values; It is easy to see that such formulas 
can be converted to formulas over Boolean variables, and 
that each clause can be converted to many clauses each operating only on 
three variables. All these details are none of our concern; 
The main issue, which we will try to mimic in the quantum case, 
 is that the history of a Turing machine can be verified 
locally.

\subsection{The Quantum analog- Sketch}
The idea of the quantum proof is very similar. 
We know that $L$ is in $QMA$; 
Thus, there exists a quantum circuit, using two-qubit gates, 
which accepts an input $x$ with some witness $|\xi\ra$ with high probability 
(we will assume it is exponentially close to $1$) if $x$ is in $L$ and rejects 
with exponentially close to one probability if $x \not\in L$ given 
any witness. 
We want to reduce this problem to the local Hamiltonian problem, 
i.e. to construct a Hamiltonian which will have small eigenvalue in the 
$x\in L$ case and only large eigenvalues otherwise. 

How to construct the analog? 
Drawing from the Cook-Levin proof, we want the history of the computation 
to be our witness, which we hope to be able to verify locally.  
Our first guess for the quantum witness would thus be 
the sequence of states which constitute the history of the computation: 
\begin{equation}
|x\ra|\xi\ra, U_1|x\ra|\xi\ra, U_2U_1|x\ra|\xi\ra, ..., U_T\cdots U_2U_1|x\ra|\xi\ra.\end{equation}
However, there is a serious problem with this suggestion. 
Let us assume for a moment that $U_1$ is simply the identity gate, and all 
we want to check is  whether the first and second states given to us by 
the prover are the same, and we want to do this via a local Hamiltonian. 
In general, we want to design a local Hamiltonian which
when applied on $|\alpha\ra|\alpha\ra$ it behaves differently than 
when applied on   $|\alpha\ra|\beta\ra$, if $|\beta\ra$ is quite different 
from  $|\alpha\ra$. 
The problem is that a local Hamiltonian has access only to
 the reduced density matrix 
of the state $|\alpha\ra|\beta\ra$ to five qubits at a time. 
In other words, 
$\la \eta|H|\eta\ra$ for any state $|\eta\ra$ 
will be exactly the same if we move to $|\eta'\ra$ as long as 
it has the same reduced density matrices as $|\eta\ra$ on all 
sets of five qubits.
It is very easy to construct two states which agree on all density 
matrices of five qubits, but are completely different due to their overall 
correlations or entanglement. 
Hence, using only local Hamiltonians we cannot hope to be able to verify 
the correctness of the time evolution if the states are given to us 
sequentially. However, entanglement which was the source of this problem, 
can also help us solve it. Consider the following superposition
\begin{equation}
\frac{1}{\sqrt{2}}(|0\ra|\alpha\ra+|1\ra|\beta\ra)
\end{equation}
From the reduced density matrix of just the {\it first} qubit, we can 
learn a lot about whether the states $|\alpha\ra$ and $|\beta\ra$
are the same or different; in fact, the reduced density matrix of
the first qubit tells us the angle between these two states, as one can 
easily verify. This means that if the histories are given to us in 
{\it superposition}, there is hope that local measurements or observables like our local Hamiltonian will be able to verify the correctness of the time 
evolution. 

The idea is therefore to ask the prover for the history 
of the computation, not in the form of sequential states but rather 
in a superposition over all {\it time leafs}:
\begin{equation}
|\eta\ra=\frac{1}{\sqrt{T+1}}\sum_{t=0}^T U_t....U_1|x,\xi\ra|t\ra
\end{equation}

we will see later how this state can actually be verified
for correctness. 
In the book\cite{kitaev} this idea of moving 
from time evolution to a time-independent local Hamiltonian is attributed to 
Feynman\cite{fey}.

Except for this main difference of using superposition over time 
instead of sequential time, there is another essential difference 
in the proof. 
In the classical case
the eigenvalues are integers, and so to show soundness one only has to
 show that the resulting formula is not satisfiable if $x$ is not 
in the language. 
The corresponding statement would be that the smallest eigenvalue 
of $H$ is not $0$; In the classical case, this automatically means 
that it is at least $1$. In the quantum case, due to the continuous 
nature of the model, the fact that the smallest eigenvalue is larger 
than $0$ is not enough; One actually has to show that it is at least 
polynomially bounded away from zero, because the accuracy achieved 
by the verification process is only polynomial, i.e. we can only 
amplify a polynomial separation and not an exponentially small 
separation.  To bound the lowest eigenvalue from below 
Kitaev uses a geometrical argument, augmented with some nice 
ideas of how to perform the analysis involving the known theory of 
random walks on the line, represented here by the time axis.  

\subsection{The reduction}
Let $L$ be a problem in QMA. Then there exists a quantum circuit $Q$ 
with two-qubit gates 
$U_1, \dots , U_T$ such that for an input $\ket{x}$ and
 a witness $\ket{\xi}$ 
the output qubit has more than $1-e^{-n}$ probability to collapse on
 $\ket{1}$ if $x \in L$ and less than $e^{-n}$ probability to 
collapse the first qubit on $\ket{0}$ otherwise. Given this sequence of gates, 
we will construct an input to the local Hamiltonian problem, 
i.e. a sequence of local matrices.  
 For now, our matrices 
will not be completely local, but instead will operate on 
two qubits among the $n$ computer qubits plus an extra 
$T+1$ dimensional Hilbert space, which will serve as a clock, and which is augmented to the right of all the other qubits. 
We will modify the Hamiltonian later on so it is truly $5$-local, 
but for the sake of simplicity we present the main part of the
 proof using this extra $T+1$
 dimensional 
Hilbert space. We denote by the subscript $C$ the subspace of the 
clock. 
The subscript $i$ at the foot of operators means that they operate on qubit 
number $i$; The projection operator $\Pi^{|\alpha\ra}$ means project 
on the subspace spanned by $|\alpha\ra$. 
\ignore{For abbreviation, 
we use $\Pi^0$, $\Pi^1$, instead of $\Pi^{|0\ra}$, $\Pi^{|1\ra}$.} 
 Our Hamiltonian will be a sum of three main terms, 
 $H = H_{in} + H_{out} + H_{prop}$. 
\begin{itemize}
\item $H_{in}$ is a matrix that checks that the input for the first $n$ qubits is indeed $x$, where we do not care about the witness; It can be anything. 
This check need to verify that the $i$th bit is indeed $x_i$, at time $0$, 
for all $i$ between $1$ to $n$. This is done  
by projecting the state to time $0$ (by projecting the clock state to time $0$), 
and then projecting the 
remaining state to the space orthogonal to $|x_i\ra$: 
\begin{equation}
H_{in}=\sum_{i=1}^n \Pi^{|\neg x_i\ra}_i\otimes |0\ra\la 0|_C
\end{equation}
\item$H_{out}$ is a matrix that checks that the output is $1$
at time $T$, again, by first projecting the clock to time $T$ and then projecting the state to the subspace orthogonal to $|1\ra$ on the first qubit
which carries the answer of the quantum circuit: 
\begin{equation}
 H_{out} = \Pi^{|0\ra}_1\otimes |T\ra\la T|_C
\end{equation}
\item$H_{prop}$ checks that the propagation of the computational 
process is done according to the given circuit.
It is a sum of $T$ terms,\begin{equation}
 H_{prop} =  \sum_{t=1}^T H_{prop}(t)
\end{equation} 
where each term checks that the propagation from time 
$t-1$ to $t$ is correct: 
\begin{equation}\label{prop}
H_{prop}(t)=
\frac{1}{2} (I\otimes \ket{t}\bra{t} + I\otimes \ket{t-1}\bra{t-1}-
U_t \otimes \ket{t}\bra{t-1} - U_t^\dag \otimes \ket{t-1}\bra{t})
\end{equation}
\end{itemize}

During the proof of the completeness part, 
it will become clear why each of these terms really verifies 
what we claim it does. Note that each term in the above
 Hamiltonian indeed satisfies 
our constraints of being Hermitian, positive semi-definite and of norm at most $1$; There is one problem, which is that it is only $log$-local and 
not local, since it operates on 
two qubits among the main qubits plus the clock space 
(which can be represented by logarithmically many qubits, which is the reason why we call it $log$-local.) We will fix this problem only much later
and for now we work with the Hamiltonian $H$ as defined.  
 To complete the reduction 
 we also need to specify $a$ and $b$; 
We let $a=1/T^{10}$, and $b=1/4(T+1)^3$. 

The claim is that the constructed Hamiltonian has an eigenvalue less than $a$ 
if $x$ is in the language that the quantum circuit accepts, 
 and otherwise all the eigenvalues of the Hamiltonian are larger than $b$.  
Once we prove both claims (completeness and soundness) we will be done; 
The two together imply that solving the local Hamiltonian problem 
for the Hamiltonian that is associated with a certain circuit, 
is a way to decide the answer of the circuit, (i.e. 
solving the Hamiltonian problem is QMA hard: any QMA problem 
can be solved using a machine that solves the local Hamiltonian problem.)
It will remain only to deal with the locality problem which we will do
at the very end.

\subsection{Completeness}
To prove completeness,  we want to show that
 a ``yes'' instance of the QMA problem 
transforms to a ``yes'' instance in the Local Hamiltonian problem. 
If $x\in L$, the $H$ we constructed has an eigenvalue smaller than $a$. 
For this, it suffices to prove the following claim: 
\begin{claim}
If $x$ is accepted by the circuit $Q$, for some quantum witness
$|\xi\ra$, with probability which is larger than $1-\epsilon$, 
 then the Hamiltonian $H=H_{in}+H_{prop}+H_{out}$
 constructed above given $x$ and the circuit $Q$
has an eigenvector with eigenvalue $\le \epsilon$.
\end{claim} 
\begin{proof}
To see why the claim is true, 
in analogy with the classical case, the state we will use 
is the history of the computation
 \begin{equation}
\ket{\eta} 
= \frac{1}{\sqrt{T+1}}\sum_{t=0}^{T} U_t U_{t-1} \dots U_1 \ket{\gamma_0} 
\otimes \ket{t}\end{equation}
where $\ket{\gamma_0}$ is the state at the beginning of the computation 
(a tensor product of the input and the witness to the machine) and 
$\ket{t}$ is a clock state. 

The intuition is  that this state is ``almost'' a zero
eigenstate of the  Hamiltonian $H$, since is ``almost'' 
satisfies all the tests 
this local Hamiltonian checks.  
More formally, we claim that 
\begin{equation}
\la \eta|H|\eta\ra\le \epsilon.
\end{equation}
which suffices to prove the claim. 

To calculate $\la \eta|H|\eta\ra$ 
we first note that 
\begin{equation}
H_{in}|\eta\ra=0.
\end{equation} 

It is less obvious but can be easily checked that for each $t=1,...T$ 
\begin{equation}
H_{prop}(t)|\eta\ra=0.
\end{equation}

The reader is recommended to verify this step, since it explains the 
definition of the propagation Hamiltonian, which is one of the main
conceptual steps in the proof. The intuition
 is that the propagation Hamiltonian is composed 
of four parts, all confined to the projections on the span of the two 
time leafs $|t-1\ra$ and $|t\ra$. Two terms in the Hamiltonian $H_{prop}(t)$,  
 $I\otimes|t\ra\la t|+I\otimes|t-1\ra\la t-1|$  correspond simply 
to picking out the state at those times. In addition, there are two 
extra terms: the term, $U_t\otimes|t\ra\la t-1|$ 
which corresponds to a forward propagation in time, 
and a term $U_t^\dagger\otimes|t-1\ra\la t|$ 
which corresponds to backwards propagation in time; 
When applied on the projection of the state to the two time steps 
$|t-1\ra$ and $|t\ra$, the forward propagation in time term picks just the 
 $t-1$ time step and propagates it forward by applying $U_t$ to it,  
and then the resulting state gets canceled 
with the $t$ time step; The same happens with the backwards propagation term
which picks up the time step $t$, propagates it one step backwards 
by applying $U_t^\dagger$ and then this term gets  canceled  with it.  
with the $t-1$ time step.

It is left to check what happens to $|\eta\ra$ when we apply $H_{out}$. 
When we apply $H_{out}$ on $|\eta\ra$ we get 
a projection on the part of $|\eta\ra$ which rejects. 
Since the probability for rejection is $\le\epsilon$, 
we get that the norm squared of 
$H_{out}|\eta\ra$ is at most $\epsilon$, and hence 
 $\la \eta|H_{out}|\eta\ra=\|H_{out}|\eta\ra\|^2\le \epsilon.$ 
Hence, the minimal eigenvalue of $H$ is less than $\epsilon$.
\end{proof}

\subsection{Soundness}
To complete the reduction, we need to show that if $x \notin L$, 
the minimal eigenvalue of $H$ is larger than the chosen $b$. 
\begin{theorem}\label{sound}
\label{p2}If $x \notin L$ then the minimal eigenvalue of $H$ 
is $\ge\frac{1}{4(T+1)^3}$.
\end{theorem}
\begin{proof}
To prove this theorem we will put together 
several lemmas. The idea is to write $H$ as a sum of two Hamiltonians, 
 $H_1 = H_{in} + H_{out},~ H_2 = H_{prop}$,
and to use the following geometrical lemma, 
which gives a lower bound on the lowest eigenvalue of 
a sum of two Hamiltonians, given some conditions 
on the eigenvalues and eigenspaces of the two Hamiltonians. 

\begin{lemma}
\label{l3}Let $H_1$ and $H_2$ be two Hermitian
positive semi-definite matrices, 
and let $N_1$ and $N_2$ be the eigenspaces of the eigenvalue $0$, 
respectively.
 If the angle between $N_1$ and $N_2$ is some $\theta > 0$, and the 
second eigenvalue of both $H_1$ and $H_2$ is $\geq \lambda$ then 
the minimal eigenvalue of $H_1 + H_2 \geq \lambda \sin^2 (\theta/2)$. 
\end{lemma}
\begin{proof} Consider an eigenvector of $H_1+H_2$,
 $\ket{\delta}$  such that $\|\ket{\delta}\| = 1$.
 For at least one of the subspaces $N_1$ or $N_2$ 
The angle between $\ket{\delta}$ and this subspace is at least 
 $\frac{\theta}{2}$. W.L.O.G let this subspace be $N_1$. 
We have 
\[ \bra{\delta}(H_1 + H_2) \ket{\delta}= 
\bra{\delta}H_1\ket{\delta} +\bra{\delta} H_2 \ket{\delta}\ge
 \bra{\delta}H_1\ket{\delta}.
\]
 We write 
\[\ket{\delta}=\ket{\mu}+\ket{\mu^\perp}\]
where $\ket{\mu},\ket{\mu^\perp}$ are the projections 
of $\ket{\delta}$ onto $N_1$ and the orthogonal subspace to $N_1$ 
respectively. 
Then 
\[
\bra{\delta}H_1\ket{\delta}=\bra{\mu^\perp}H_1\ket{\mu^\perp}\ge \|\mu^\perp\ra\|^2\lambda\]
where the first equality follows from the fact that 
$N_1$ and and its complement are invariant to the application of $H_1$
and the second follows from the definition of $H_1$ and $\lambda$. 
We also know that 
 $ \|\mu^\perp\ra\|^2\ge sin^2(\theta/2)$ because the angle between 
$N_1$ and $|\delta\ra$ is at least $\theta/2$, and this completes 
the proof. \end{proof}

To use the geometrical lemma, we will assume $x \notin L$ and 
 give lower bounds on the second 
eigenvalues of $H_1$ and $H_2$, as well as a lower bound on
$\theta$.
We will first bound the second eigenvalues of $H_1$ and $H_2$. 

\begin{lemma}
The second eigenvalue of $H_1$ is at least $1$. 
\end{lemma}

\begin{proof} 
The second eigenvalue of $H_1$ is $\geq 1$ since $H_{in}$ and $H_{out}$ 
are projections and hence their eigenvalues are $0$ and $1$. 
Since the eigenspaces of the eigenvalue $1$ of $H_{in}$ and $H_{out}$ 
are orthogonal (because they operate on different times), they commute, and so their second eigenvalue is simply the minimal second eigenvalue of the two.
\end{proof}

\begin{lemma}
The second eigenvalue of $H_2=H_{prop}$ is at least $\frac{1}{2(T+1)^2}$. 
\end{lemma}

\begin{proof}
It turns out that
for this and further arguments it 
is simpler to look at $H_{prop}$ in a rotated basis. 
The eigenvalues of a matrix are not changed when looked 
at in a different basis. 
Hence we define the unitary matrix $R$ as follows: 
\begin{equation}R=\sum_{t=0}^T U_{t}...U_1 \otimes |t\ra\la t|.
\end{equation}
$R$ is unitary since it is a block diagonal matrix with each of its 
blocks unitary. 
What $R$ does is basically rotate the basis in each time leaf 
to the basis which one gets if one applies the first $t$ 
computation steps on the computational basis.
Hence, in the new rotated basis, the computation is simply the identity. 
 Now, it is easy to check that 
 \begin{equation}R^\dagger H_{prop}(t)R=  \frac{1}{2}(I \otimes \ket{t}\bra{t} 
+ I \otimes \ket{t-1}\bra{t-1} - I \otimes \ket{t-1}\bra{t} 
- I \otimes \ket{t-1}\bra{t})\end{equation}
We can write $H_{prop}=I\otimes A$ where $A$ is a $(T+1)\times (T+1)$ of the form: 
\begin{displaymath}
A=     \left ( \begin{array}{cccccccc}        
                \frac{1}{2}&-\frac{1}{2}&0&0&0&0&0&0\\
                -\frac{1}{2}&1&-\frac{1}{2}&0&0&0&0&0\\
                0&-\frac{1}{2}&1&-\frac{1}{2}&0&0&0&0\\
                0&0&-\frac{1}{2}&1&-\frac{1}{2}&0&0&0\\
                0&0&0&-\frac{1}{2}&1&-\frac{1}{2}&0&0\\
                0&0&0&0&-\frac{1}{2}&1&-\frac{1}{2}&0\\
                0&0&0&0&0&-\frac{1}{2}&1&-\frac{1}{2}\\
                0&0&0&0&0&0&-\frac{1}{2}&\frac{1}{2}\\
               \end{array}
\right )
\end{displaymath}

\begin{displaymath}
=    I- \left ( \begin{array}{cccccccc}        
                \frac{1}{2}&\frac{1}{2}&0&0&0&0&0&0\\
                \frac{1}{2}&0&\frac{1}{2}&0&0&0&0&0\\
                0&\frac{1}{2}&0&\frac{1}{2}&0&0&0&0\\
                0&0&\frac{1}{2}&0&\frac{1}{2}&0&0&0\\
                0&0&0&\frac{1}{2}&0&\frac{1}{2}&0&0\\
                0&0&0&0&\frac{1}{2}&0&\frac{1}{2}&0\\
                0&0&0&0&0&\frac{1}{2}&0&\frac{1}{2}\\
                0&0&0&0&0&0&\frac{1}{2}&\frac{1}{2}\\
               \end{array}
\right )=I-B
\end{displaymath}

The eigenvalues of $R^\dagger H_{prop}R$ 
or equivalently of $H_{prop}$ are simply the eigenvalues 
of $A$ (with multiple appearances), or $1$ minus those of $B$; 
It suffices then to find the eigenvalues of $B$. 

Interestingly, the matrix $B$ is a familiar matrix
from the theory of random walks and we will use this fact 
in the analysis of its eigenvalues. 
For a direct proof see Kiteav\cite{kitaevbook}. 
Here we refer to 
the theory of random walks due to its intriguing connection with 
the subject at hand. For a nice exposition of random walks, see 
Lovasz's survey\cite{walks}.
Returning to our matrix $B$,  
it turns out that it is the stochastic matrix 
corresponding to a simple random walk on  the time axis, from $0$ to $T$. 
 with a loop at both ends. 
The largest eigenvalue of this matrix is $1$,
corresponding to the eigenvector which is the 
 uniform limiting distribution. This eigenvalue 
gives the $0$ eigenvalue of $A$ and hence of $H_{prop}$.
In random walk theory, 
one is very interested in the second eigenvalue of the stochastic matrices 
corresponding to random walks since the second eigenvalue
 is directly related  
to the rate at which the random walk mixes to its limiting distribution. 
$B$'s second largest eigenvalue $\lambda_2$ is bounded from below by
the conductance $\phi$ of the graph on which the random walk is applied,  
using Jerrum and Sinclair's bound\cite{sinclair}: 
 \begin{equation}1-\lambda_2\ge \phi^2/2\end{equation}
The conductance of the random walk is $\frac{1}{T+1}$ which gives 
$1-\lambda_2\ge \frac{1}{2(T+1)^2}. $
Since $1$ minus the second largest eigenvalue of $B$
is exactly the second smallest eigenvalue of $A$,  
this implies the desired result.  
\end{proof}

It is left to give a lower bound on the angle between the two null spaces. 
 
\begin{lemma}\label{trig}
The angle between $N_1$ and $N_2$ satisfies 
$sin^2(\theta/2)\ge \frac{1}{2(T+1)}$.
\end{lemma}

\begin{proof}
$H_1=H_{in}+H_{out}$ is a projection, and hence the null 
space is simply the subspace orthogonal to the space on which 
$H_1$ projects. Hence, $N_1$ is equal to 
the direct sum of three subspaces: 
\begin{equation}N_1=
 (\ket{x}\bra{x}\otimes W \otimes \ket{0}\bra{0})\oplus
 (\ket{1}\bra{1} \otimes W \otimes \ket{T}\bra{T})\oplus_{t=1}^{T-1}
(W \otimes \ket{t}\bra{t})\end{equation}
where $W$ is the entire Hilbert space for the remaining of the qubits.  
$N_2$, the null space of $H_{prop}$, 
 is exactly the space spanned by
 all valid computations starting with an arbitrary state $|\alpha\ra$ 
on the qubits of the input and witness together. These are all 
states of the form:
\begin{equation}\label{eta}\ket{\eta} =
\frac{1}{\sqrt{T+1}} \sum_{t=0}^T U_t \dots U_1 \ket{\alpha}\otimes \ket{t}.\end{equation}
The fact that such states are in the null space of $H_{prop}$ was 
shown before; The fact that all states in the null space of $H_{prop}$ are 
of this form follows from looking at the rotated $R^\dagger H_{prop} R$, 
as before. 
The null space of the rotated $H_{prop}$ is simply the entire space 
on the computer register times the null space of 
the clock matrix $A$; 
 The  null space of the matrix $A$ is exactly all constant vectors.
This is a standard claim, following from the fact that the random 
walk $B$ defines on the line is aperiodic, ergodic, and 
 converges to the uniform vector. (One can readily prove this fact also 
from scratch, by considering the effect of $A$ on the eigenvector 
corresponding to eigenvalue $1$,  
and looking at the maximal coordinate.) 
\ignore{ 
Let $f$ be a function on the time axis. 
$A$ applied on this function outputs a new function. 
 the value of which at a given point is the original value minus
the average of the original values 
of on both sides. This is true except for the edges, 
where the value is simply the original value minus the value 
at the neighbor. 
If $v$ is an eigenvector with eigenvalue zero, then its real and imaginary parts are also eigenvector with eigenvalue zero. 
Hence, consider only real vectors. 
Consider $v_i$ a coordinate with the maximal value of $v$. 
If this coordinate is not one of the edge points, then 
since $v$ is a zero eigenvector, it follows from applying $A$ 
that $v_i$ equals the average of $v_{i-1}$ and $v_{i+1}$, but 
since $v_i$ is maximal, then both $v_{i-1}$ and $v_{i+1}$
are equal to $v_i$, and hence they are maximal too. 
In this way we prove that all points are constant. 
If the maximum is achieved on the edge, say $v_0$, 
it follows from applying $A$ 
that $v_0$ is equal to $v_1$ and hence $v_1$ is maximal,
 and we are back to the previous argument. }

Now, to find $N_2$, the null space of $H_{prop}$,  
we have to rotate the null space of $RH_{prop}R^\dagger$ 
(which is all the Hilbert space on the computer qubits times constant
vectors on the clock space) back to the original basis, by applying 
$R$ and $R^\dagger$ from both sides. 
It is easy to see that for any state 
\begin{equation}|\alpha\ra\otimes \frac{1}{\sqrt{T+1}}\sum_{t=0}^T |t\ra\end{equation}
rotating it back gives a state of the form 
of equation \ref{eta}.

We now want to bound the angle between $N_1$ and $N_2$, which is the 
minimal angle between two vectors from both spaces. 
Any vector in $N_2$ is of the form of  equation \ref{eta}, i.e. a history
of a certain computation, 
and the angle $\phi$ between such a history $|\eta\ra$ 
and $N_1$ is given by 
\begin{equation}
cos^2(\phi)=\|\Pi_{N_1}|\eta\ra\|^2
\end{equation}
where $\Pi_{N_1}$ denotes the projection onto $N_1$. This is true since 
$|\eta\ra$ is of norm $1$.
Thus we have 
that the angle $\theta$ between $N_1$ and $N_2$ 
is the minimal angle $\phi$ between a history vector   and the space $N_1$,  or equivalently:  
\begin{equation}\label{above}
cos^2(\theta)=max_{|\eta\ra\in N_2}\{\|\Pi_{N_1}|\eta\ra\|^2\}
\end{equation}
We now claim that for any $|\eta\ra\in N_2$ we have: 
\begin{equation}\label{upp}
\|\Pi_{N_1}|\eta\ra\|^2 \le 1-\frac{1}{2(T+1)}.
\end{equation}
The proof of this will complete the proof of the lemma,
 using equation \ref{above}.  
To prove the upper bound of equation \ref{upp}, we observe that 
the norm squared of the projection onto $N_1$ is simply the sum 
of the norms squared on the projections on the 
different parts of $N_1$, as a direct sum of subspaces. 
If we write $N_1$ as a direct sum of the different spaces 
spanned by 
times $t = 1, ... ,T-1$, then since $|\eta\ra$ 
 is the uniform superposition over time, the projection of 
 $\ket{\eta}$ on each of the middle time step  gives 
 $\frac{1}{T+1}$, and so the total contribution 
of the middle time leafs is $\frac{T-1}{T+1}$. 

We now claim that the contribution of the first and last leafs 
together is far from the maximal possible contribution $\frac{2}{T+1}$. 
Intuitively, this is due to the fact that the projection on $N_1$ sums up the 
projection on $x$ as input in the beginning of the computation plus 
the projection on ``accept'' at  the end of the computation. 
However, since  $|\eta\ra$ represents a valid computation by a circuit 
that does not accept $x$, it cannot be the case that both 
projections are maximal. 
To quantify this statement, we 
observe that we can write $|\eta\ra$ as a sum of two states: 
\begin{equation}|\eta\ra=\frac{1}{\sqrt{T+1}}\sum_{t=0}^T |\gamma_t\ra\otimes|t\ra=
\frac{1}{\sqrt{T+1}}\sum_{t=0}^T (a|\gamma^1_t\ra+b|\gamma^2_t\ra)\otimes|t\ra=a|\eta_1\ra+b|\eta_2\ra\end{equation}
where $|\gamma^1_0\ra$ is the normalized projection of $|\gamma_0\ra$ on the 
input being $x$, and $|\gamma^2_0\ra$ is the normalized
 projection on the orthogonal 
subspace, and $|\gamma^1_t\ra$,$|\gamma^2_t\ra$ are simply the states 
obtained from the initial states by applying the computation. 
The norm sqaured of the projection of the first time leaf of 
 $|\eta\ra$, $\frac{1}{\sqrt{T+1}}|\gamma_0\ra\otimes |0\ra$, onto $N_1$ is $\frac{a^2}{T+1}$. 
The norm squared of the projection of the last time leaf of 
$|\eta\ra$, $\frac{1}{\sqrt{T+1}}|\gamma_0\ra\otimes |T\ra$,
 onto $N_1$ is the norm squared of 
\begin{equation}\frac{1}{\sqrt{T+1}}(a|\delta^1_T\ra+b|\delta^2_T\ra)
\end{equation}
where $|\delta^1_T\ra,|\delta^2_T\ra$ are the projections of 
$|\gamma^1_T\ra,|\gamma^2_T\ra$ on ``accept'', respectively 
(we are using the linearity of projection.)
Now we have that 
\begin{equation}\||\delta^1_T\ra\|^2\le e^{-n}\end{equation}
since the circuit accepts with probability less than $e^{-n}$ 
if $x$ is not in the language. 
Hence, 
\begin{equation}
\|a|\delta^1_T\ra+b|\delta^2_T\ra\|\le 2e^{-n}+b^2\end{equation}
and so the total norm squared of the projection on $N_1$ 
is at most 
\begin{equation}\|\Pi_{N_1}|\eta\ra\|^2 
\le\frac{1}{T+1}(a^2+2e^{-n}+b^2)+\frac{T-1}{T+1}\le (1-\frac{1}{2(T+1)})
\end{equation}
using the fact that $a^2+b^2=1$. This completes the proof.\end{proof}

To complete the proof of the theorem \ref{sound}, 
we simply apply the geometrical 
lemma \ref{l3} using the bounds we have shown for the eigenvalues 
and for $\theta$. \end{proof}

This completes the proof of the hardness of Local Hamiltonian for 
QMA, if we are allowed to use Hamiltonians which operate on 
spaces of polynomial dimension; 
In the next section we make the last step that is needed to convert 
the Hamiltonian to a $5$-local Hamiltonian consisting of terms 
operating on five qubits only. 

\ignore{
Since $x \notin L$, either the projection of ${\eta}$ on 
$\ket{x}\bra{x}\ket{0}\bra{0}$ is $0$, or the projection on 
$\ket{1}\bra{1}\ket{l}\bra{l}$ is $\leq \frac{1}{3}$ (it is not
$0$ since in QMA we allow for an error of $\frac{1}{3}$). 
Therefore times $t=0$ and $t=l$ together give at most $\frac{4}{3(l+1)}$. 
We get that $cos^2\alpha \leq 1 - \frac{2}{3(l+1)}$.}

\section{Improving from $log$ local to $5-$local}
To move from 
operators on the entire clock to local operators, 
represent the time in  unary representation on $T$ qubits which will serve 
as the clock qubits. For example, time $t=4$ 
is represented by the $T$ qubit state $|111100\dots00\ra.$ 
To modify the Hamiltonian accordingly, we replace all operators on the clock 
space by operators that
 operate on three qubits at most. We apply the following 
modifications: 
\begin{eqnarray}
&&\ket{t}\bra{t-1} ~~~~~\longmapsto \ket{110}\bra{100} \otimes I\\\nonumber
&&\ket{t-1}\bra{t}~~~~~\longmapsto \ket{100}\bra{110} \otimes I\\\nonumber
&&\ket{t}\bra{t}~~~~~~~~~~\longmapsto \ket{110}\bra{110} \otimes I\\\nonumber
&&\ket{t-1}\bra{t-1}\longmapsto \ket{100}\bra{100} \otimes I
\end{eqnarray}
 where in all these cases $\ket{110}\bra{100}$ or the similar terms
 operate on 
qubits $t-1, t, t+1$ of the clock qubits and the identity $I$ operates
 on the remaining $T-2$ clock qubits.
This will hold in all terms of $H_{prop}$, 
except for two exceptions to the above - when $t=1$ and $t=T$, in order not 
to refer to bits $0$ and $T+1$ of the clock which do not exist.
For $t=1$ we will drop the first bit of the 3-bit operator, 
so the operator $\ket{1}\bra{0}$ on the original
 clock becomes $\ket{10}\bra{00}$ on the first two bits; 
and similarly $\ket{0}\bra{1}$ on the original clock 
 becomes $\ket{00}\bra{10}$ on the first two bits in the unary clock. 
For the case $t=T$ we drop the $3^{rd}$ bit of 
the operators in the same manner.
The final $H_{prop}$ which we get is  
\begin{equation}
H'_{prop}(t)=
\frac{1}{2} (I\otimes \ket{110}\bra{110} + I\otimes \ket{100}\bra{100}-
U_t \otimes \ket{110}\bra{100} - U_t^\dag \otimes \ket{100}\bra{110})
\end{equation}
where the three qubit operators operate on qubits $t-1,t,t+1$. 
For $H'_{prop}(1)$,  $H'_{prop}(T)$ we get a slightly different 
expression with the clock operators operating only on two qubits 
as explained above. 

As for $H_{in}$ and $H_{out}$, we again change $|t\ra\la t|$ to be 
an operator on three qubits for the middle time leafs and two qubits 
for the beginning and end leafs. 

We first claim that restricting ourselves to the subspace 
spanned by states with the clock qubits being in
 valid unary representations, all previous claims hold. 
Explicitly, as can be easily checked:   
\begin{claim}
For any state $\ket{\eta}$ which represents a valid history of a computation,
(in unary representation)  
we still get $ H'_{prop} \ket{\eta} = 0$.
\end{claim}
From this, using exactly the same arguments as used before, 
we have that 
\begin{claim}
If $|\eta\ra$ is a history of an accepting computation, 
then $\la \eta|H'|\eta\ra\le \epsilon.$
\end{claim}
Hence, completeness will go through with these modifications. 
However, soundness will not go through because of the following reason. 
The $T$ qubits that we have introduced have many more possible states 
except for valid unary
representations of some time step. The $H'$ we have defined operates on such 
states as well; To prove soundness, 
we need to show that among such states there are no 
states of small eigenvalue in case of $x$ not in the language. 
This might be complicated, and we resort to a different solution.

In addition to the modifications of the existing terms in the Hamiltonian, 
we introduce a new term which penalizes 
the state of the clock qubits if they are not unary representation
of some time step. 
We call this term $H_{clock}$; It locally checks that the clock 
bits are a valid unary representation. 
All we need to check is that two consequent bits cannot be in the state  $|01\ra$; This can be done by a sum of local projections, as follows:
\begin{equation}
 H'_{clock} = \sum_{t=1}^T \ket{01}\bra{01}_{t-1,t} \otimes I. 
\end{equation}
Our (truly!) final Hamiltonian is defined to be 
\begin{equation}
H'=H'_{in}+H'_{out}+H'_{prop}+H'_{clock}
\end{equation}
Clearly, $\ket{\eta}$ (with a unary clock) is an eigenvector of eigenvalue 
$\le \epsilon$  
of the new Hamiltonian $H'$, since it is a zero eigenvector of $H'_{clock}$, 
and so completeness is preserved. 

For the proof of soundness, we observe that $H'$ 
keeps the subspace that is
 spanned by all states in which the clock 
qubits are valid unary representations invariant; 
Let us call this subspace   ${\cal D}$. 
The orthogonal subspace,  
${\cal D}^\perp$, is also invariant under the operation of $H'$. 
$H'$ operates on ${\cal D}$ just as the previous $H$ did, and hence on this 
subspace the lower bound on the eigenvalues holds as before; 
On the orthogonal subspace  ${\cal D}^\perp$ the eigenvalue of $H'$ 
is at least $1$ since $H'_{clock}$ detects at least one violation.  
Hence, overall, the lower bound from theorem \ref{sound} holds here too. 
This completes the proof of completeness of $5$-local Hamiltonian. 

\begin{remark}\underline{{\bf Why 5?}} 
We remark here regarding the necessity of three qubits Hamiltonians 
instead of one qubit Hamiltonians to control the propagation in time. 
One can naively suggest to use the one qubit Hamiltonian $|1\ra\la 0|$ 
operating on the $t^{th}$ clock qubit to represent the propagation from 
$t-1$ to $t$, instead of the three qubit operator we use.
This suggestion does not work for the following reason: 
the history states will no longer be eigenstates of $H'_{prop}$, 
since the one qubit time propagation terms might cause valid 
time leafs to propagate to invalid ones; E.g., $|11100000\ra$ will 
propagate by $U_6\otimes|1\ra\la 0|_6$ to 
$|11100100\ra$. This does not happen when three qubit operators are used.
It is an open question whether this obstacle can be overcome to show that 
$3$-local or $4$-local Hamiltonian is QMA complete; 
See open question \ref{less}.  
\end{remark}

\section{Discussion and Open Questions}
We have  presented here a beautiful result by Kitaev which
we believe is a fundamental stepping stone for the field 
of quantum complexity. We collect here a list of open questions 
it raises.

The first set of problems is related to the 
question of the expressiveness of the class QMA. 
There are hundreds of NP complete problems, from an enormous 
variety of fields; 
So far, the only interesting quantum MA complete problem we know of 
is the Local Hamiltonian problem.  

\begin{open}\label{more}
Find more quantum MA complete problems. 
\end{open}

\noindent In particular, 
 
\begin{open}
Is there a natural QMA complete problem which is not quantum related? 
\end{open}

We have proved that $5$-local Hamiltonian is $QMA$ complete. 
What is the importance of the number five? 
It is unknown whether $5$ qubits are necessary.  Perhaps 
 even $2-$locality suffices to achieve 
{\cal QMA} completeness.   
This is very different from the classical situation, where 
it is known that $3$-SAT is NP complete but $2$-SAT 
can be solved in polynomial time. 
\begin{open}\label{less}
What is the complexity of $k$-local Hamiltonian with $k=2,3,4$? 
\end{open}

The next question is related to 
the definition of the class $QMA$:  

\begin{open}\label{two}
Is {\cal QMA} with two sided errors the same as {\cal QMA} with one sided error? 
\end{open}

This holds in the classical case, and the question is whether
it holds quantumly.  Kitaev and Watrous\cite{kitwat} show the equivalence of 
one and two sided errors in certain cases (quantum interactive proofs with more rounds) but the proof does not carry over to this case. 

Another open question which is related to the definition 
of $QMA$ is 

\begin{open}\label{same}
Is {\cal QCMA}$=${\cal QMA}?  
\end{open}

Due to the results presented in this survey, it seems 
reasonable to assume that the answer is yes. 
Our intuition behind this conjecture is that 
the quantum verifier limits itself in its tests of the quantum state 
to the reduced density matrices of five qubits; 
It therefore does not care about longer range entanglement.
Perhaps such states that are specified by short range entanglement 
can be efficiently generated. 
In other words: 
\begin{open}  
Consider a (possibly very complicated) $n$ qubit state $|\xi\ra$. 
Is there an efficient circuit that generates a state $|\xi'\ra$
which has (almost) the same reduced density matrices to any subset of five qubits? 
\end{open}

If this can be done, this will prove the equality $QCMA=QMA$ since 
 the classical witness can be 
the description of the quantum circuit, and the verifier can generate 
the state on its own.
A proof that shows this equivalence is likely to be 
very insightful regarding quantum correlations. 
This question touches upon the interesting question 
of whether it is possible to develop a quantum analog 
 of the beautiful theory of pseudo-random generators;
 In pseudo-random generation, one generates  probability
 distributions that are very different from the uniform 
distribution
but such that any circuit of some restricted set (say of bounded size)
cannot tell the difference. 
The question we are asking is of a similar type, and can be viewed as 
a {\it pseudo quantum generator} type question,  since we need the state 
 to pass the test of the very restricted verifier who only looks at 
sets of five qubits. 

The results presented here highlight a very interesting 
connection between Hamiltonians and unitary gates or quantum circuits, 
which Kitaev attributes to Feynman\cite{fey}. 
We view this connection 
as fascinating and potentially very powerful. In particular,
  in exactly the same way  $QMA$ circuits are translated to a local
 Hamiltonian, 
one can also translate $BQP$ circuits to a local Hamiltonian; 
It is very interesting to ask whether the other direction of 
moving from groundstates of local Hamiltonians to efficient circuits that 
generate them also holds. We cannot hope for a constructive version 
of this direction since it is $QMA$ hard, 
 but an existence proof of such short circuits for ground states of local Hamiltonians  will be very interesting, leading to a positive answer to open question \ref{same} and to implications to the
 very important task of quantum state generation (see \cite{amnon}.) 
\begin{open}
Given a local Hamiltonian, does there exist a polynomial size quantum circuit 
that generates (a state with non negligible projection on) its ground state?  
\end{open}

We remark that Feynman's point of view\cite{fey}
of moving from circuits to time independent Hamiltonians, 
enables one to translate short computation time into large 
spectral gap of Hamiltonians. The spectral gap of $\Omega(1/T^3)$ 
achieved in kitaev's proof is
 indeed expressed in terms of the computational time $T$. 
 It is interesting to compare this  
to the framework of adiabatic quantum computation\cite{farhi} 
in which the opposite direction 
of moving from large spectral gaps to efficient state generation is taken, 
when one aims at generating a groundstate of a final Hamiltonian $H_f$ by 
designing a sequence of local Hamiltonians $H_0,...H_f$ 
all with large spectral gaps.   

We end with a more general open problem which seems related to 
the results presented here:  
\begin{open}
Does the quantum analog of the PCP theorem hold?  
Can we prove hardness of approximation for quantum computation?   
\end{open}

Hopefully, the results and open questions presented in this survey 
will provide easy access to research in what we view as a fascinating area. 
\section{Acknowledgements}
One of us (D.A.) wishes to thank Avi Wigderson for insightful comments
during Kitaev's first lecture about the subject\cite{kitaev}.

\end{document}